\documentstyle[12pt]{article}
\begin{document}
\large
\renewcommand {\baselinestretch} {1.0}
\setcounter {page} {1}
\begin{center}
{\bf Some Consequences of the Law of Local Energy Conservation
in Electromagnetic field}
\vspace{0.2cm}
\par
Kh. M. Beshtoev
\vspace{0.2cm}
\par
Joint Institute for Nuclear Research, Joliot Curie 6,
141980 Dubna, Moscow region, Russia
\vspace{0.3cm}
\end{center}
\par
{\bf Abstract}\\

\par
At electromagnetic interactions of particles there arises  defect
of masses, i.e. the energy is liberated since the particles of the different
charges are attracted. It is shown that this change of the effective mass
of a particle in the external electrical field (of a nucleus) results in
displacement of atomic levels of electrons. The expressions describing
these velocity changes and displacement of energy levels of electrons
in the atom are obtained.

\section{Introduction}

At a gravitational interaction of particles and bodies a defect masses
arises [1], i.e. there
an energy yield appear since the bodies (or particles) are attracted.
In the previous work [1] it was shown, that the radiation spectrum
(or energy levels) of atoms (or nuclei) in
the gravitational field has a red shift since the effective mass of
radiating electrons (or nucleons) changes in this field.
This red shift is equal to the red shift of the radiation spectrum
in the gravitational field measured in existing experiments. The same
shift must arise when the photon (or $ \gamma $ quantum) is passing
through the gravitational field if it participates in gravitational
interactions. The absence of the double effect in the experiments
means that photons (or $ \gamma $ quanta) are passing through the
gravitational field without interactions.
\par
In work [2] it was shown that changing of the effective
mass of a body (or a particle) leads to changing of velocity and length
measurement units (relative to standard measurement units). An expression
describing the advance of the perihelion of the planet (the Mercury) is
obtained. This expression is mathematically identical to Einstein's
equation [3] for the advance of the perihelion of the Mercury but in a flat
space. The same situation must take place at the electromagnetic interaction
of particles and nuclei.
\par
This work is devoted to search for influence of attractive electromagnetic
interaction on atomic levels of electrons.

\section{\bf Some Consequences of the Law of Local Energy Conservation
in Gravitational field}

a). Let us consider the influence of the  external electrical field
$\varphi = -e\frac{Z}{r}$ ($Z e$ is electrical charge of the external field,
$r$--distance) on characteristics of an electron (particle) having
a small velocity. The law of local energy conservation in the classical case
has the following form:
$$
E = \frac{m v^2_1}{2} + e \varphi_1 = \frac{m v^2_2}{2} + e \varphi_2
\eqno(1)
$$
or
$$
\frac{m (v_1^2 - v^2_2)}{2} = Z e^2 (\frac{1}{r_1} - \frac{1}{r_2})
\eqno(2)
$$
The Eqs. (1) and (2) characterize the balance between kinetic and potential
energies (the smaller the energy--the bigger another energy and back).
\par
In a more strict form the law of local energy conservation can be rewritten
in the form
$$
E = m c^2 + \frac{m v^2_1}{2} + e \varphi_1 =
    m c^2 + \frac{m v^2_2}{2} + e \varphi_2
\eqno(3)
$$
Then the Eq. (3) can be rewritten in th following form:
$$
E = m c^2 (1 + e \frac{\varphi_1}{m c^2}) + \frac{m v^2_1}{2}  =
    m c^2 (1 + e \frac{\varphi_2}{m c^2}) + \frac{m v^2_2}{2} .
\eqno(4)
$$
After introduction the new masses are

$$
m' = m (1 + e \frac{\varphi_1}{m c^2}) \qquad
m'' = m (1 + e \frac{\varphi_2}{m c^2})
\eqno(5)
$$
and new velocities--
$$
v_1'^2 = \frac{v^2_1}{(1 + e \frac{\varphi_1}{m c^2})} \qquad
v_2'^2 = \frac{v^2_2}{(1 + e \frac{\varphi_2}{m c^2})} .
\eqno(5')
$$
Then Eq. (4) acquires the following form:
$$
E = m' c^2 + \frac{m' v'^2_1}{2}  =
    m'' c^2 + \frac{m'' v'^2_2}{2} .
\eqno(6)
$$
The eq. (6) means that in an external electrical field the effective mass of
the electron (particle) changes. For clarification of this question let us
consider a body (or particle) with mass $m$ in the external electrical
field $\varphi$ in point $r$ and write the law of local energy conservation
for this system
$$
m c^2 = E = m c^2 + \frac{m v^2}{2} + e \varphi_ =
    m' c^2 + \frac{m' v'^2}{2} ,
\eqno(7)
$$
where
$$
m' = m(1 + e \frac{\varphi}{m c^2}) \qquad \Delta m = m - m' =
- e \frac{\varphi}{m c^2} ,
$$
and
$$
\delta v^2 = v^2 - v'^2 =  v^2 (\frac{e \varphi}{m c^2}) \qquad or \qquad
\frac{\Delta v^2}{v^2} =  \frac{e \varphi}{m c^2} .
\eqno(8)
$$
Eqs. (7), (8) means that changing the mass ($\Delta m c^2$) of the electron
(particle) in the external electrical field goes on the kinetic energy of
this electron (particle).
\par
To which result do come? In contrast to the classical physics the velocity
of the electron (particle) is $v'$ and
$$
\Delta v^2 = v^2 - v'^2 =  v^2 \frac{e \varphi}{m c^2}, \qquad
\Delta v =  v \frac{e \varphi}{2 m c^2} .
\eqno(9)
$$
\par
The following question arises: how can we see this changing of the
effective mass or velocity?
\par
This effect, in principle, is very small. We can register this effect
in atomic transitions as changing of the atomic levels relatively the standard
levels. It is clear, that we must learn these transitions in nuclei with
large $Z$ where this effect will be sufficiently visible.
\par
The atomic levels are given by the following expression [4]:
$$
E_n = \frac {\alpha^2 m _ {eff} c^2} {2} \frac {Z^2} {n^2} [1 + \frac {\alpha^2
Z^2} {n} [\frac {1} {(j + 1/2)} - \frac {3} {4n}] +...]  ,
\eqno (10)
$$
where
$$
\alpha = \frac {e^2} {4 \pi \hbar c}; \qquad n ' = 0, 1, 2...;
$$
$$
j = \frac {1} {2}, \frac {3} {2}, \frac {5} {2}...; \qquad
n = n ' + j + \frac {1} {2} = 1, 2, 3... \qquad ,
$$
$n$ is an orbital number.
\par
The law of the local energy conservation is fulfilled if we take into
account that the mass of the electron $m' = m_{eff}$ in the connected
(bound) state is
$$
m_{eff} = m - 2 E_n ,
\eqno(11)
$$
where $E_n$ is a radiating energy and $\frac{m v^2}{2} \simeq E_n$ is the
energy of the orbital movement of the electron.
\par
Putting (11) into (10) we come to the following expression:
$$
E_n = \frac {\alpha^2 (m - 2 E_n/c^2) c^2} {2} \frac {Z^2} {n^2} [1 + \frac {\alpha^2
Z^2} {n} [\frac {1} {(j + 1/2)} - \frac {3} {4n}] +...]  ,
\eqno (12)
$$
From (12) we come to the following expression for the real electron levels
$E_n$ for the nucleus with charge $Z$:
$$
E_n = \frac {\alpha^2 m c^2} {2} \frac {Z^2} {n^2} [1 + \frac {\alpha^2
Z^2} {n} [\frac {1} {(j + 1/2)} - \frac {3} {4n}] +...] \cdot
$$
$$
\frac{1}{1 +
 {\alpha^2} \frac {Z^2} {n^2 } [1 + \frac {\alpha^2
Z^2} {n} [\frac {1} {(j + 1/2)} - \frac {3} {4n}] +...]}  ,
\eqno (13)
$$
It is clear that electron in the connected (bound) state cannot  radiate
a photon if it's mass does not change.
\par
When we take into account the law of energy conservation in a strict form
changing the energy of the electron levels $\Delta E_n$  is
$$
\Delta E_n
 = - \frac {\alpha^2 m c^2} {2} \frac {Z^2} {n^2} [1 + \frac {\alpha^2
Z^2} {n} [\frac {1} {(j + 1/2)} - \frac {3} {4n}] +...]  \cdot
$$
$$
[ {\alpha^2} \frac {Z^2} {n^2} [1 + \frac {\alpha^2
Z^2} {n} [\frac {1} {(j + 1/2)} - \frac {3} {4n}] +...]]  ,
\eqno (14)
$$
If the transitions take place between levels $n$ and $m$, then in the
expression (14) it is necessary to do the following changing:
$$
\frac{1}{n^2} \to (\frac{1}{n^2} - \frac{1}{m^2})
\eqno(15)
$$

\section{Conclusion}

\par
At electromagnetic interactions of particles there arises defect
of masses, i.e. the energy is liberated since the particles of the different
charges are attracted. It is shown that this change of the effective mass
of a particle in the external electrical field (of a nucleus) results in
displacement of atomic levels of electrons. The expressions describing
these velocity changes and displacement of energy levels of electrons
in the atom are obtained.
\par
It is necessary to stress that the same effects will take place in the
nuclei at the strong interactions of the protons and neutrons.\\

\par
{\bf References}

\par
\noindent
1. Kh.M. Beshtoev, JINR Commun. P4-2000-45, Dubna, 2000;
\par
quanta-ph/0004074.
\par
\noindent
2. Kh.M. Beshtoev, JINR Commun. P4-2001-107, Dubna, 2001;
\par
astro-ph/0107430.
\par
\noindent
3. A. Einstein, Ann. Phys., 1911, v.35, p.898;
\par
M. Born, Einstein's Theory of Relativity, Dover, N.Y., 1962.
\par
\noindent
4. S.S. Schweber, An Introduction to Relat. Quantum Field Theory,
\par
(Row-Peterson and Co., N. Y., 1961).

\end{document}